\documentclass[12pt,preprint]{aastex}
\newcommand{\be}{\begin{equation}}
\newcommand{\ee}{\end{equation}}
\def\ergs{{\rm\,erg\,s^{-1}}}
\newcommand{\msun}{{\rm M}_{\sun}}

\slugcomment{Accepted by ApJ}

\begin{document}

\title{The hard X-ray spectral evolution in X-ray binaries and
its application to constrain the black hole mass of ultraluminous
X-ray sources}

\author{Qingwen Wu$^{1}$ \& Minfeng Gu$^{2,3}$}

\altaffiltext{1}{International Center for Astrophysics, Korean
Astronomy and Space Science Institute, Daejeon 305348, Republic of
Korean; Email: qwwu@shao.ac.cn}

\altaffiltext{2}{Shanghai Astronomical Observatory, Chinese Academy
of Sciences, Shanghai, 200030  China}

\altaffiltext{3}{Joint Institute for Galaxy and Cosmology (JOINGC)
of SHAO and USTC, 80 Randan Road, Shanghai 200030, China}

\begin{abstract}
    We investigate the relationship between the hard X-ray photon
    index $\Gamma$ and the Eddington ratio ($\xi=L_{X}(0.5-25\ \rm keV)/L_{Edd}$)
    in six X-ray binaries (XRBs) with well constrained black hole masses and
    distances. We find that different XRBs follow different
    anti-correlations between $\Gamma$ and $\xi$ when $\xi$ is less than a critical
    value, while $\Gamma$ and $\xi$ generally follow the same positive
    correlation when $\xi$ is larger than the critical
    value. The anti-correlation and the positive correlation may suggest that
    they are in different accretion modes (e.g., radiatively
    inefficient accretion flow (RIAF) and standard disk).
    We fit both correlations with the linear least-square method for individual sources,
    from which the crosspoint of two fitted lines is obtained.
    Although the anti-correlation varies from source to source, the crosspoints
    of all sources roughly converge to the same point with small scatter
  ($\log \xi=-2.1\pm0.2, \Gamma=1.5\pm 0.1$), which may correspond to the transition point between RIAF and standard
  accretion disk. Motivated by the observational evidence for the similarity of the X-ray spectral evolution of
    ultraluminous X-ray sources (ULXs) to that of XRBs, we then constrain the black hole masses for seven
    ULXs assuming that their X-ray spectral evolution is similar to that of XRBs.
     We find that the BH masses of these seven luminous ULXs are around
    $10^{4}\msun$, which are typical intermediate-mass BHs (IMBHs).
    Our results are generally consistent with the BH masses constrained from the timing
    properties (e.g., break frequency) or the model fitting with a multi-color disk.
\end{abstract}


\section{Introduction}
  Ultraluminous X-ray sources (ULXs) are pointlike, nonnuclear X-ray
  sources with X-ray luminosities between $10^{39}$ and $10^{41} \ergs$, well in excess of
  the Eddington limit for a stellar mass black hole (BH, $L_{\rm Edd}=1.3\times10^{38}(M_{\rm BH}/\msun) \ergs$, with
  $M_{\rm BH}$ the mass of the BH). The existence
  of temporal variability on a variety of timescales in these ULXs suggests that they are accreting objects.
  However, the true nature of these sources is still unclear, especially their BH
  mass (for recent reviews, see Miller \& Colbert 2004; Fabbiano \& White 2006). Current models for these objects center around two
  alternatives: (1) they are ``intermediate-mass BHs" (IMBHs) with
  mass $M_{\rm BH}\simeq 10^{2}-10^{5} \msun$, accreting at
  sub-Eddington rates (e.g., Colbert \& Mushotzky 1999; Makishima et al.
  2000; Kaaret et al. 2003; Miller et al. 2003; Yuan et al. 2007; Liu \& Di Stefano 2008); (2) they are stellar mass BHs, accreting with
  super-Eddington rates (e.g., Watarai et al.
  2001; Begelman 2002; Vierdayanti, et al. 2006), or X-ray binaries (XRBs) with anisotropic
   emission (King et al. 2001), or beamed XRBs with relativistic jets
  directly pointing toward us (e.g., Mirabel \& Rodriguez 1999).
  In several ULX systems (e.g., NGC 1313 X-2, M81 X-9, etc), detection
  of emission nebulae surrounding the ULX supports isotropic
  emission from the central source, which cannot be explained
  through beaming (Pakull \& Mirioni 2003).

    Spectral state transitions are essential characteristics of XRBs
    and both the low/hard state and high/soft states have been
    observed in most XRBs. Typically, the X-ray energy spectrum in the
   hard state can be well described as a power law with a photon index
   $\Gamma=1.5-2.1$. The energy spectrum in soft state can be
   described with a thermal disk component and a power law with a photon index
   $\Gamma=2.1-4.8$. Some sources also show the steep power law
   state with $\Gamma>2.4$, which differs from the soft state in
   that the power law component, rather than the disk component, is
   dominant (for a review see Remillard \& McClintock 2006).
    The first two types of spectra (low/hard state and high/soft state) are naturally
    explained by the existence of two different
   stable accretion flows, with a hot, optically thin, geometrically
   thick radiatively inefficient accretion flow (RIAF, e.g., Narayan \& Yi 1994;
    Blandford \& Begelman 1999; Narayan et al. 2000),
   which can exist only at low luminosities, as well as a cool,
   optically thick and geometrically thin standard disk (Shakura \&
   Sunyaev1973). However, the physical origin of the very high state
   is still an open problem, and possible mechanisms include inverse Compton scattering
   as the operant radiation mechanism (e.g., Zdziarski \& Gierli$\acute{\rm n}$ski 2004) or bulk motion
   Comptonization in a converging sub-Keplerian flow near the BH (e.g., Titarchuk \& Shrader 2002).
   The hard X-ray photon index is found to be related to
   luminosity (or Eddington ratio) in both XRBs and active
   galactic nuclei (AGNs). In particular, the photon index is
   anti-correlated with the Eddington ratio in both low/hard
   state XRBs (e.g., Kalemci et al. 2005; Yamaoka et al. 2005; Yuan et al. 2007) and low
   luminosity AGNs (LLAGNs, Gu \& Cao 2008) when the Eddington ratio
   is less than a critical value.
   However, a positive correlation between the photon index and the Eddington
   ratio is found in XRBs with higher Eddington ratio (e.g., XTE J1550-564, Kubota \& Makishima 2004)
   and luminous AGNs (e.g., Wang et al. 2004; Shemmer et al. 2006).

     The X-ray spectra of ULXs have been extensively studied with
   $XMM-Newton$ and $Chandra$. The spectra of some ULXs can
   be explained as radiation from a standard disk with an
   additional high-energy power law spectrum (e.g., Colbert \& Mushotzky 1999;
    Miller et al. 2003; Miller et al. 2004), which is also the
   standard accretion model for high/soft state XRBs and luminous
   AGNs. The spectra of some other ULXs can be represented by a single power law, suggesting
   a similarity to the low/hard state of XRBs and low luminosity
   AGNs (e.g., Colbert et al. 2004). The long-term variations in the X-ray spectra of some
   ULXs also show the low/hard to high/soft state transition or high/soft
   to low/hard state transition, reminiscent of the behavior
   of classic XRBs (e.g., Kubota et al. 2001; Roberts et al. 2004; Feng \& Karret
   2006; Soria et al. 2006). Moreover, the recent observation of a break frequency
   and a quasi-periodic oscillation (QPO) in the power density
   spectrum of the well-known ULXs (e.g., M82 X-1 and NGC 5408 X-1) also suggests
   the similarities between ULXs and XRBs (e.g., Dewangan et al. 2006; Strohmayer et al. 2007).

   In this work, we revisit the X-ray spectral evolution in some well
   studied XRBs, which is then applied to constrain the BH masses of several ULXs assuming that their X-ray
    spectral evolution is similar to that of XRBs and AGNs.
    In $\S\S 2$ and 3, we introduce the sample and our results, respectively. The
    discussion and summary are presented in $\S\S 4$ and 5,
    respectively.

\section{Sample}

    We search the literatures for XRBs observed
    by the $Rossi\ X-ray\ Timing\ Explorer\ (RXTE)$. To explore the X-ray spectral evolution, we focus on the XRBs
    with well determined hard X-ray spectral slopes during the state transition, in addition to
     well determined BH masses and
    distances. The X-ray data of all the selected XRBs were observed during the decay of the
    outburst, i.e., transition from soft to hard state.
    Table 1 shows the list of the XRBs we analyze in this
    paper, which includes seven outbursts of six XRBs.

    To utilize the X-ray spectral evolution of XRBs to constrain the black hole masses for ULXs, the multiple X-ray data
    are also collected from the literature for ULXs. In this paper, we only tentatively
    include seven luminous sources with the 0.3-10 keV X-ray luminosities near or larger than $10^{40}
    \ergs$.  All the sources have multiple (at least three epochs) high resolution
     $XMM-Newton$ and/or $Chandra$ observations.
    Most of the X-ray data have been fitted by several models (e.g.,
    simple absorbed power-law model (PL), multicolor disk blackbody
    model plus power-law model (MCD+PL), etc.), and the photon index and luminosity of best fittings
    were used in this work (Table 2).
    In addition, the peak luminosity and spectral slope of M82 X-1 measured with $RXTE$ during its flare, is also
    included, since it is less contaminated by the host galaxy.
    The sample of ULXs is shown in Table 2.

\section{Result}

   \subsection{X-ray spectral evolution in XRBs}
   In Fig. 1, we present the relation between the photon
  index, $\Gamma$, and the Eddington ratio, $\xi$, of the XRB sample,
   where the photon index $\Gamma$ is fitted in the 3-25 keV band.
   We define the Eddington ratio $\xi=L_{X}(0.5-25\ \rm keV)/\it L_{\rm Edd}$, in which $L_{X}(0.5-25\ \rm keV)$
  is extrapolated from the unabsorbed 3-25 keV band luminosity, since which was regarded
   as more represent the bolometric luminosity than the pure hard X-ray
   luminosity. It can be clearly seen that
  the hard X-ray photon index $\Gamma$ is strongly correlated with the
  Eddington ratio $\xi$ when $-2\lesssim\log \xi\lesssim0$, even through this relation becomes shallower at higher
  Eddington ratios for some sources. In the low Eddington ratio (e.g., $\log \xi\lesssim
  -2$) part, the anti-correlation is generally present, which actually has already been found in many
  occasions (e.g., Yamaoka et al. 2005; Yuan et al. 2007). However, the
  most remarkable result in our work is that different sources follow
  the different anti-correlations (Fig. 1). It is also interesting
  that the photon index $\Gamma$ is roughly constant in H1743-222
  when $-2.5\lesssim \log \xi \lesssim -2$.

  We fit the anti-correlation points between the Eddington ratio and the photon index with linear least-square
  method for each source. We also fit all datapoints at the positive correlation ($-2\lesssim \log \xi \lesssim
  -1$) region as a whole with the same method, since all sources roughly follow the same
  correlation (short dashed line in Fig.
  1). Since the correlation becomes flatter, and a large scatter is present at higher Eddington ratios
   (i.e., $\log \xi \gtrsim-1$), these data points are not
  included in the fit. The source H1743-222 is
  not included in the fitting due to the large scatter of photon index $\Gamma$ in anti-correlation part.
  Consequently, for each source, the crosspoint between the positive and anti-correlation part
   can be obtained through the fitted linear relation on both parts.
  Although the anti-correlation varies from source to source, we
  find the crosspoints of all XRBs roughly converge to the same point with small scatter
  ($\log \xi=-2.1\pm0.2, \Gamma=1.5\pm 0.1$).  For comparison with AGNs, we also plot the best
  fits to the relation between the photon index and the Eddington
  ratio in LLAGNs (long dashed-line in Fig. 1, Gu \& Cao 2008) and
  luminous AGNs (long dashed-line in Fig. 1, Shemmer et al. 2006),
  where the LLAGNs include 27 LINERs and 28 local Seyfert galaxies,
  while the luminous AGNs include 26 radio-quiet QSOs. We also
  include one radio quiet Seyfert galaxy NGC 4051, which show strong
  X-ray spectrum and flux variation (green solid circles in Fig. 1). The X-ray spectral photon
  indices and fluxs (seven epoch observations) are from Lamer et al. (2003), and
  $3\times10^{5}\msun$ BH mass is adopted in our work (McHardy et al. 2004).
  We find that the photon indices are positive correlate to the Eddington ratios
  for NGC 4051, which is similar to
  that of QSOs. In doing so, we have extrapolated the X-ray luminosities to
  0.5-25 keV band for all AGNs. It is interesting that the anti-correlation in LLAGNs and
  the positive correlation in luminous AGNs cross at point ($\log \xi=-2.78, \Gamma=1.61$), where the Eddington
  ratio $\xi$ is several times less than that of XRBs. The positive correlation of NGC 4051 and anti-correlation of
   LLAGN sample cross at point ($\log \xi=-2.52, \Gamma=1.55$).
  Therefore, it seems that the hard X-ray spectral
  evolution in XRBs and AGNs can be formulized as:
  \be
  \Gamma = \kappa (\xi+2.1^{+0.7}_{-0.2}) +1.5^{+0.1}_{-0.1}
  \ee
where $\kappa$ is the slope and $\xi$ is the Eddington ratio. Keep
in mind
  that the difference of the crosspoint between AGNs and XRBs is mainly caused by the different bolometric
  correction factors (see discussions in paragraph 4 of Section 4).

 \subsection{Constraints on the BH masses of luminous ULXs}

   Similar to XRBs, it has been found that many ULXs also show strong X-ray spectral
    evolution. In particular, it is found that the X-ray photon index of NGC 1313 X-1 is
    correlated with luminosity, while these two parameters are anti-correlated in NGC 1313
    X-2, which are similar to that of both XRBs and
    AGNs  (Feng \& Kaaret 2006). This phenomenon has also been found in many other
    ULXs (e.g., Roberts et al. 2004). The similarity between ULXs, XRBs and AGNs implies that
    there may be similar physics behind the phenomena, which motivates us to explore
    the properties of ULXs utilizing the unification of ULXs and XRBs (and/or AGNs).
    Specifically in this work, the BH masses of ULXs can be constrained if we
    assume that the BH accretion is scale-free and that their X-ray spectral evolution is similar to
    that of XRBs and AGNs. Assuming the spectral evolution can be described by Equation 1,
    we thus calculate the BH masses of ULXs through the
    least-squares linear fit on the available data for each source.
    The spectral evolution of ULXs is shown in Fig. 2.
    We note that all available data of most ULXs belong to either the positive
    correlation part or the anti-correlation part. This results in a unique estimate of BH mass for each ULX.
    In the case of NGC 1313 X-2, we only use the anti-correlation data points in calculating the BH mass,
    since there are only a few data points in the positive correlation
    part. The spectra of NGC 5408 X-1 are
    similar to that of XRBs in very high state, we only derive the upper limit of its BH mass since
    the relation between photon index and Eddington ratio of XRBs
    becomes shallower at higher Eddington ratios (see Fig. 1).
    Our results show that all the BHs in
    these luminous ULXs are IMBHs of around $10^{4}\msun$ (see Table 2).
    It should be noted that the dearth observational data for most of
    the ULXs precludes an estimate of the uncertainty of BH mass.

  The shape of the fluctuation power spectral density (PSD, e.g., QPO, and/or
break frequency) of ULXs is potentially a very powerful tool to
study the central engine of ULXs. In particular, the characteristic
frequency of breaks in the PSD slope can be used to infer BH masses
based on a direct scaling of properties between XRBs and AGNs (see
McHardy et al. 2006). The putative break frequency has been detected
in three ULXs (NGC 5408 X-1: Soria et al. 2004; M82 X-1: Dewangan et
al. 2006; NGC 4559 X-7: Cropper et al. 2004). We calculate the BH
masses of these three ULXs (see Table 2) based on the $\nu_{\rm
break}-M_{\rm BH}-L_{ \rm Bol}$ relation (see McHardy et al. 2006).
To utilize this relation, we estimate the bolometric luminosity from
0.1-200 keV band for these three ULXs, which was extrapolated from
the unabsorbed 0.3-10 keV luminosity. We find that the BH masses
constrained from the X-ray spectral evolution are consistent with
those derived from the break frequency within a factor of two except
for NGC 5408 X-1, for which only the upper limit is given. However,
the BH mass of NGC 5408 X-1 derived from the spectral evolution
method in this work is roughly consistent with that derived from the
multi-color disk fittings and QPO (Strohmayer et al. 2007)(see also
Table 2). This might suggest that it is not appropriate to constrain
the BH mass from the break frequency for very high state of sources
(e.g., NGC 5408 X-1), which is also the case in AGNs with near or
super Eddington accretion (e.g., McHardy et al. 2006).

\section{Discussion}

  In this paper we investigate the relation between the hard X-ray
photon index $\Gamma$ and the Eddington ratio $\xi$ in a small
sample of BH XRBs. The Eddington ratio varies by several orders of
magnitude from $10^{-5}$ to 1, and the spectra also show strong
evolution. We find that different XRBs follow different
anti-correlations between $\Gamma$ and $\xi$ when $\log \xi\lesssim
-2$, while a positive correlation is present when $-2\lesssim\log
\xi\lesssim 0$ (Fig. 1).
 The anti-correlation and the positive correlation have been found in both XRBs
(e.g., Kalemci et al. 2005; Yamaoka et al. 2005; Yuan et al. 2007)
and AGNs (e.g., Wang et al. 2004; Shemmer et al. 2006; Gu \& Cao
2008). The qualitative behavior of the anti-correlation is
consistent with the predictions of the spectrum produced from the
RIAF model (see Fig. 3a of Esin et al. 1997), and the jet
contribution in the X-ray waveband may not be important in these
XRBs (e.g., Yuan \& Cui 2005; Wu et al. 2007). The Comptonization of
thermal synchrotron and bremsstrahlung photons from the hot
electrons in the RIAF is the dominated cooling mechanism at low
Eddington ratios. As the accretion rate increases, the optical depth
of the RIAF also increases, increasing the Compton $y$-parameter,
thereby leading to a harder X-ray spectrum (detailed calculation
will be presented in Yuan, F. et al. 2008, Inpreparation).
 However, the physical reason for the positive correlation in both XRBs and luminous AGNs
(e.g., QSO) when $\log \xi\gtrsim -2$ is still unclear. One
possibility is that the Comptonization of thermal blackbody photons
from the outer thin disk will be the dominant cooling mechanism
(e.g., Narayan 2005), since the standard thin disk will transit to
optically thin, hot accretion flows (e.g., RIAF or transition layer
model in Titarchuk et al. 1998) at certain radii. When the
luminosity or accretion rate starts to increase, the transition
radius decreases, then the outer thin disk radiation becomes
stronger and Compton cooling by disk photons in the inner hot
accretion flows becomes more efficient, and the electron temperature
goes down, thereby leading to a softer X-ray spectrum (e.g.,
Zdziarski et al. 2000). This can also naturally explain the possible
positive correlation between luminosity, hard X-ray photon index and
low frequency QPO, if we assume that the low frequency QPO is
related to the transition radius of two accretion flows (e.g.,
Titarchuk \& Fiorito 2004). Another possibility is that the standard
disk is not truncated, and extends to the innermost stable circular
orbit (ISCO). As the accretion rate increases, the fraction of
accreting energy released to the corona decreases (e.g., Merloni \&
Fabian 2002; Liu et al. 2002; Wang et al. 2004), the corona becomes
weak and shrinks, the optical depth of the corona decreases,
reducing the y-parameter, and thus leading to a softer X-ray
spectrum. It is still unclear what's physical reason leading to the
constant photon index in H1743-222 when $-2.5\lesssim \log\xi
\lesssim -2$ (Fig. 1).

It is interesting to note that the traditional hard state can be
divided into two parts, i.e., bright hard state and faint hard
state, by the crosspoint ($\log \xi \sim -2.1$) of the
anti-correlation and the positive correlation. The X-ray spectral
evolution of the bright hard state with $\log \xi > -2.1$ is the
same as that of high/soft state of XRBs and luminous QSOs. It
implies that accretion process in this bright hard state may be the
standard accretion disk as that of the high/soft state. The
observations of both the cool accretion disk component and the
relativistic broad Fe K emission line in the bright hard state of GX
339-4 (e.g., Miller et al. 2006) and some bright Seyfert I galaxies
( Nandra et al. 2007; for recent review, Miller 2007) give strong
supports to this scenario. In contrast, the RIAF model becomes
important in the faint hard state of the XRBs with $\log \xi
\lesssim -2.1$. If this is the case, the anti-correlation and the
positive correlation may suggest they are in RIAF mode and standard
accretion disk mode, respectively, and the crosspoint may correspond
to the accretion mode transition, which, however, is different from
the popular idea that the disk transition occurs at $\Gamma=2.1$
(i.e., high/soft to low/hard state transition).

The physical reason for different anti-correlations between the
photon index and the Eddington ratio in different sources is
unclear. One possibility is that the differences are caused by the
different transitional condition of RIAF and SSD (e.g., electron/ion
temperature etc.) since the outer boundary conditions of RIAF play
an important role in determining its structure and spectrum (e.g.,
Yuan et al. 2000). The electron temperature on the outer boundary
will be lower if RIAF and SSD transit at higher Eddington ratios
since electrons will cool more efficiently at higher Eddington
ratios. The electron temperature on the boundary decreases, the
electron temperature in the inner region will also decrease,
reducing the Compton $y$-parameter, thereby leading to a softer
X-ray spectrum. This is consistent with Fig. 1, in which the steeper
X-ray spectral evolution corresponds to the higher Eddington ratio
of the crosspoint between the anti-correlation and the positive
correlation (e.g., XTE J1748-288). Normally, the range of the
transition point is small since the electron temperature will
decrease abruptly when plasma density increase to a critical value
(e.g., GX 339-4 in Miyakawa et al. 2007). Consequently, the
transition of RIAF and SSD of different sources can occur at
different Eddington ratios, however the difference is rather small,
which leads to the same crosspoint with small scatter for our XRBs
in Fig. 1.

 One should be cautious to compare the crosspoint of the AGNs
 to that of XRBs (Fig. 1), since the accretion process
 in some bright Seyferts in the LLAGN sample of Gu \&
 Cao (2008) may be also similar to that of QSOs (see discussions
  in paragraph 2 of section 4). However, this kind of
 bright Seyferts only constitute a small
 fraction of the LLAGN sample, therefore, the crosspoint of AGNs will
 not change a lot. The second point we should note is that the bolometric
 correction factor ($f_{\rm cor}=L_{\rm Bol}/L_{\rm X} (0.5-25\rm keV)$) of XRBs should
 be several times lower than that of AGNs, which will also lead to the
  crosspoint of AGNs being lower by several times if we assume that
  the change of the correlation between $\xi$ and $\Gamma$ is caused
   by the ratio of the \emph{total bolometric} luminosity and the Eddington luminosity.
   We find that the bolometric correction factor $f_{\rm cor}$ of XRBs is roughly 2-5 times lower than that
  of AGNs, where the bolometric luminosity of XRBs is estimated from the 0.1-200keV band X-ray luminosity and
  the bolometric luminosity of AGNs is estimated from multi-wavebands observations (e.g., Elvis et al. 1994; Ho 1999).
  This is consistent with our results that the Eddtington ratio of the crosspoint of AGNs
  is several times less than that of XRBs (Fig. 1). The bolometric correction
  factors only vary several times for different kind sources (e.g., XRBs, ULXs, and AGNs), and will not affect our main
  conclusions.

   It is commonly accepted that ULXs are accreting objects,
   however, their nature is still unclear, of which the mass of
   the central BH is one of the most important issues. On the
   assumption of the same X-ray spectral evolution between ULXs and
   XRBs (and/or AGNs), which is supported by various observations (e.g., Roberts et al. 2004; Winter et al. 2006;
   Feng \& Kaaret 2006; Soria et al. 2007), we suggest that the BH
   masses of these luminous ULXs are around $10^{4}\msun$, which
   are the typical value of IMBHs.
   There are uncertainties in the estimated BH
   masses of ULXs, which originate from several aspects. Firstly, it is partly from the uncertainties in
   determining the crosspoint of XRBs, which are caused by the limited
   data, the uncertainties on distances and BH masses determination and also the possible small
   intrinsic transition range. Secondly, the limited datapoints of ULXs
   used in linear least-square fit may cause uncertainties. There are only several
   observations for most of the ULXs in this sample, except NGC 1313 X-1
   and X-2, therefore, more uniform observations using high resolution and high
   sensitivity X-ray telescopes are desired to further constrain
    their BH masses.
    Most importantly, more complete and higher
   quality X-ray data of XRBs are needed to constrain their X-ray
    spectral evolution, and to test the reality of a universal
   crosspoint in XRBs. The validity of our method can be tested by comparing the estimated BH
   masses
   with those derived from independent methods. In addition to the consistency with the BH
   masses
   derived using $\nu_{\rm break}-M_{\rm BH}-L_{ \rm Bol}$ relation
   for XRBs and AGNs (McHardy et al. 2006), we find that the BH masses of M81 X-9 and NGC 1313
   X-1 are also consistent within a factor of two with those
   constrained from the multi-color disk fittings (Miller et al.
   2004) (see Table 2). These consistency may support the validity of our
   method, which further implies that the assumption of similarity
  between ULXs and XRBs is likely reasonable. It was reported
  that some ULXs have an energy break in the 5-10keV (Stobbart et al. 2006).
  If this is the case, the bolometric luminosity would be overestimated when we using
  the $\nu_{\rm break}-M_{\rm BH}-L_{ \rm Bol}$ relation to estimate
  the BH mass , since the bolometric luminosity was derived from the 0.1-200keV luminosity, which is
  extrapolated from the unabsorbed 0.3-10keV luminosity. However, we find that
   the BH mass estimated in this way is consistent in factor of two with that
   estimated using a bolometric luminosity as 0.1-10keV luminosity.
   On the other hand, more high quality X-ray data and detailed model
   fitting is needed to further explore the energy break. In some cases,
    the contrast results can be obtained from the same data, however, with
   different model. For instance, in NGC 1313 X-1, Feng \& Kaaret (2006) find
   no evidence of an energy break based on the MCD+PL model, while the energy
   break near 5 keV was found based on the ``nonstandard model" (soft power law+hot
   accretion disk blackbody, Stobbart et al. 2006).

    We point out that the inferred BH masses of ULXs in our work
    are calculated based on the assumption that their X-ray spectral evolution is similar to that of
     the well known XRBs with luminosity less than or near Eddington luminosity,
     which is independent of the assumed radiative efficiency of the
    accretion process or specific accretion models. We do not exclude
     the possibility that these ULXs are accreting at a super Eddington
     rate with stellar mass of BHs. However, this
     needs about several thousand times of Eddington accretion
     rate to support the luminosities of $10^{41}\ergs$ observed in some ULXs
     if we assume a 10 $\msun$ BH, since both advection and photon trapping are very
     important in this case (e.g., Ohsuga et al. 2002), and we never observed XRBs
     and AGNs with such high accretion rate. Feng \& Kaaret (2007) found
     that the maximum disk color temperature, $T$, is anti-correlated with
     the disk luminosity, $L$, in NGC 1313 X-2, which is inconsistent the $L\propto T^{4}$ relation
      expected if the cool thermal component is the emission from the standard accretion
     disk as observed in XRBs. Instead, its spectral evolution can be roughly fitted
     by the $p$-free disk model with a stellar mass BH, and the $p$ value is around 0.5-0.6, where the $p$-free disk
     model is similar to the standard disk model but the radial temperature
     follows the form as $T\propto R^{p}$ and allows $p$ to
     vary with radius $R$ (Mineshige et al. 1994). Therefore,
     they suggest that the BH in NGC 1313 X-2 should be a stellar
     mass BH with super-Eddington accretion. We note that another
     possibility for NGC 1313 X-2 is that its BH is IMBH, but
      accreting through a radiatively inefficient accretion process (e.g.,
     RIAF) since its anti-correlated spectral evolution is similar to the low hard state XRBs.
     It is also difficult to explain both the positive correlation
      and the anti-correlation between the photon index and
     the Eddington ratio for all ULXs by the same $p$-free disk
     model with a stellar mass BH. What's more, the detected QPO in
     some ULXs (e.g., Dewangan et al. 2006, Strohmayer et al. 2007) contradicts the disk dominated
     $p$-free model, since the QPO is absent or weak in the disk dominated high/soft
     state of XRBs (see Remillard \& McClintock 2006, and references therein).
       Detailed model fittings of more ULXs, considering the evolution of both luminosities
     and spectra, will be our future work.

\acknowledgments We wish to thank the referee for helpful
suggestions and comments that improved the clarity of the paper. We
would like to thank X. W. Cao and F. Yuan for their valuable
discussions and constructive suggestions. Q. W. Wu thanks S.
Mineshige for his hospitality when visit Yukawa Institute for
Theoretical Physics in May, 2007, and Dr. A. Humphrey (KASI) for
help with proofreading this paper. This work is supported by the
postdoctoral fellowship of Korean Astronomy and Space Science
Institute and the National Natural Science Foundation of China under
grants 10543002 and 10703009.

\clearpage
\begin{deluxetable}{lccccc}
\tabletypesize{\scriptsize}
 \tablecaption{Data of XRBs}
 \tablewidth{0pt} \tablehead{
\colhead{Source}& \colhead{Distance (kpc)} & \colhead{BH mass
($\msun$)} & \colhead{Obs. Number} & \colhead{Outburst year} &
\colhead{Refs.$^{a}$} }

\startdata

 4U 1543-47         &   7.5    &    9.4   &  74   &  2002   & P04 \& K05 \\
 XTE J1550-564a     &   5.0    &    10.0  &  25   &  2000   & K02  \\
 XTE J1550-564b     &   5.0    &    10.0  &  12   &  2001   & K02  \\
 XTE J1118+480      &   1.8    &    6.1   &  12   &  2000   & K02   \\
 XTE J1748-288      &   8.0    &    10.0  &  23   &  1998   & R00 \\
 GX 339-4           &   8.0    &    5.8   &  6    &  1998   & K02 \\
 H1743-222          &   11.0   &    10.0  &  48   &  2003   & K06 \\

\enddata

\tablerefs{P04: Park et al. 2004; K05: Kalemci et al. 2005; K02:
Kalemci 2002; R00: Revnivtsev et al. 2000; K06: Kalemci et al. 2006}

\tablenotetext{a}{The reference of the observation, and reference of
BH mass and distance therein}


\end{deluxetable}

\clearpage
\begin{deluxetable}{lcccccccc}
\tabletypesize{\scriptsize}
 \tablecaption{Data of ULXs}
 \tablewidth{0pt} \tablehead{
\colhead{Name}& \colhead{$\Gamma$} & \colhead{Model$^{a}$}&
\colhead{$\log L_{X}$$^{b}$} & \colhead{Telescope} &
\colhead{Distance$^{c}$} & \colhead{Reference$^{d}$} & \colhead{BH
mass 1$^{e}$} & \colhead{BH mass 2$^{f}$} }

\startdata
 M81 X-9     &    $1.73$                   & MCD+PL  & 40.22  &  $XMM-Newton$  & 3.4  & M04  & $6.6\times10^{3}$  & $5^{+7}_{-2}\times10^{3}$ $^{g}$\\
 ...         &    $1.86$                   & MCD+PL  & 40.26  &  $XMM-Newton$  & ...  & M04  & ...                &...\\
 ...         &    $2.28^{+0.58}_{-0.35}$   & MCD+PL  & 40.77  &  $XMM-Newton$  & ...  & T06  & ...                &...\\
\hline
NGC 1313 X-1 &    $1.75^{+0.06}_{-0.04}$   & MCD+PL  & 40.11  &  $XMM-Newton$  & 4.13 & F06  & $7.9\times10^{3}$  &    $4^{+2}_{-1}\times10^{3}$ $^{g}$\\
...          &    $2.04^{+0.16}_{-0.15}$   & PL      & 40.19  &  $XMM-Newton$  & ...  & F06  & ...                &...\\
...          &    $2.42^{+0.07}_{-0.06}$   & PL      & 40.27  &  $XMM-Newton$  & ...  & F06  & ...                &...\\
...          &    $2.29^{+0.13}_{-0.12}$   & PL      & 40.22  &  $XMM-Newton$  & ...  & F06  & ...                &...\\
...          &    $2.16^{+0.14}_{-0.13}$   & PL      & 40.13  &  $XMM-Newton$  & ...  & F06  & ...                &...\\
...          &    $2.09^{+0.16}_{-0.19}$   & MCD+PL  & 40.20  &  $XMM-Newton$  & ...  & F06  & ...                &...\\
...          &    $2.38^{+0.08}_{-0.07}$   & PL      & 40.19  &  $XMM-Newton$  & ...  & F06  & ...                &...\\
...          &    $1.8^{+0.2}_{-0.2}$      & MCD+PL  & 40.12  &  $XMM-Newton$  & ...  & F06  & ...                &...\\
...          &    $2.39^{+0.06}_{-0.05}$   & PL      & 40.43  &  $XMM-Newton$  & ...  & F06  & ...                &...\\
...          &    $3.08^{+0.15}_{-0.14}$   & MCD+PL  & 40.11  &  $XMM-Newton$  & ...  & F06  & ...                &...\\
...          &    $1.8^{+0.14}_{-0.19}$    & MCD+PL  & 40.13  &  $XMM-Newton$  & ...  & F06  & ...                &...\\
...          &    $1.74^{+0.11}_{-0.05}$   & MCD+PL  & 40.17  &  $XMM-Newton$  & ...  & F06  & ...                &...\\
\hline
NGC 1313 X-2 &    $2.12^{+0.17}_{-0.10}$   & MCD+PL  & 39.54 &  $XMM-Newton$  & 4.13 & F06  & $2.4\times10^{4}$  &    ...\\
...          &    $1.81^{+0.12}_{-0.12}$   & PL      & 40.03 &  $XMM-Newton$  & ...  & F06  & ...                &...\\
...          &    $1.86^{+0.10}_{-0.05}$   & MCD+PL  & 40.24 &  $XMM-Newton$  & ...  & F06  & ...                &...\\
...          &    $1.71^{+0.06}_{-0.05}$   & PL      & 40.19 &  $XMM-Newton$  & ...  & F06  & ...                &...\\
...          &    $2.22^{+0.08}_{-0.07}$   & PL      & 39.73 &  $XMM-Newton$  & ...  & F06  & ...                &...\\
...          &    $2.39^{+0.08}_{-0.07}$   & PL      & 39.62 &  $XMM-Newton$  & ...  & F06  & ...                &...\\
...          &    $2.34^{+0.12}_{-0.11}$   & PL      & 39.57 &  $XMM-Newton$  & ...  & F06  & ...                &...\\
...          &    $2.50^{+0.18}_{-0.17}$   & PL      & 39.49 &  $XMM-Newton$  & ...  & F06  & ...                &...\\
...          &    $1.93^{+0.06}_{-0.05}$   & MCD+PL  & 40.42 &  $XMM-Newton$  & ...  & F06  & ...                &...\\
...          &    $1.94^{+0.24}_{-0.17}$   & MCD+PL  & 39.62 &  $XMM-Newton$  & ...  & F06  & ...                &...\\
...          &    $2.19^{+0.14}_{-0.12}$   & MCD+PL  & 39.57 &  $XMM-Newton$  & ...  & F06  & ...                &...\\
...          &    $1.90^{+0.05}_{-0.03}$   & MCD+PL  & 40.58 &  $XMM-Newton$  & ...  & F06  & ...                &...\\
\hline
M82 X-1      &    $1.67\pm0.02$            & PL      & 40.50  &  $Chandra$     & 3.63 & K06  & $6.6\times10^{3}$  & $5^{+7}_{-2}\times10^{3}$ $^{g}$\\
...          &    $1.8^{+0.11}_{-0.09}$    & PL      & 40.64  &  $XMM-Newton$  & ...  & D04  & ...                & ...\\
...          &    $2.35\pm0.15$            & PL      & 40.74  &  $RXTE$        & ...  & K06  & ...                & ...\\
\hline
NGC 5408 X-1 &    $2.61^{+0.06}_{-0.05}$   & MCD+PL  & 39.84  &  $XMM-Newton$  & 4.8  & S04  & $<4.0\times10^{3}$ & $1.5\times10^{4}$ $^{h}$\\
...          &    $2.67^{+0.18}_{-0.17}$   & MCD+PL  & 39.82  &  $XMM-Newton$  & ...  & S04  & ...                & $3.3\pm1.5\times10^{3}$ $^{i}$\\
...          &    $2.93^{+0.17}_{-0.23}$   & MCD+PL  & 39.85  &  $XMM-Newton$  & ...  & S04  & ...                & $2.5\pm1\times10^{3}$ $^{j}$\\
...          &    $2.61^{+0.18}_{-0.18}$   & MCD+PL  & 39.86  &  $XMM-Newton$  & ...  & S04  & ...                & ...\\
...          &    $2.85^{+0.16}_{-0.18}$   & MCD+PL  & 39.90  &  $XMM-Newton$  & ...  & S04  & ...                & ...\\
\hline
NGC 4559 X-7 &    $1.8\pm0.08$             & MCD+PL  & 40.38  &  $Chandra$     & 9.69 & C04  & $1.4\times10^{4}$  &    $7.6\times10^{3}$ $^{h}$\\
...          &    $2.13\pm0.08$            & MCD+PL  & 40.51  &  $Chandra$     & ...  & C04  & ...                & ...\\
...          &    $2.23^{+0.05}_{-0.04}$   & MCD+PL  & 40.33  &  $XMM-Newton$  & ...  & C04  & ...                & ...\\
\hline
NGC 4559 X-10&   $1.99^{+0.22}_{-0.21}$   & MCD+PL  & 40.16  &  $Chandra$     & 9.69 & C04  & $2.7\times10^{4}$  &    ...\\
...          &   $1.82^{+0.12}_{-0.13}$   & MCD+PL  & 40.26  &  $Chandra$     & ...  & C04  & ...  &    ...\\
...          &   $2.05\pm0.07$            & MCD+PL  & 40.14  &  $XMM-Newton$  & ...  & C04  & ...  &    ...\\
\enddata

\tablerefs{M04: Miller et al. 2004; T06: Tsunoda et al. 2006; F06:
Feng \& Kaaret 2006; K06:  Kaaret et al. 2006; D04: Dewangan et al.
2006; S04: Soria et al. 2004; C04: Cropper et al. 2004}

\tablenotetext{a}{The model used to fit the X-ray observations, PL
is the single Power Law model, and MCD+PL is the multicolor disk
(MCD) blackbody model plus Power Law model.}

\tablenotetext{b}{In unit of ergs/s. The X-ray luminosity in the
0.5-25keV band, which is extrapolated from the unabsorbed 0.3-10keV
band luminosity.}

\tablenotetext{c}{in unit of Mpc.}

\tablenotetext{d}{References for X-ray data and references for
distance and initial X-ray data therein.}

 \tablenotetext{e}{The BH mass (in unit of $\rm M_{\odot}$) estimated from spectral evolution method in this paper.}

\tablenotetext{f}{The BH mass (in unit of $\rm M_{\odot}$) estimated
from other method: ($g$) the multi-color spectrum fittings (Miller
et al. 2004); ($h$) the break frequency using $\nu_{\rm
break}-M_{\rm BH}-L_{ \rm Bol}$ relation of McHardy et al. (2006) in
this paper; ($i$) the multi-color spectrum fittings (Strohmayer et
al. 2007); ($j$) the scaling of QPO frequency assuming that the QPO
frequency is proportional to $1/\rm M_{\rm BH}$  (Strohmayer et al.
2007).}

\end{deluxetable}

\clearpage

\begin{figure}
\epsscale{1.0} \plotone{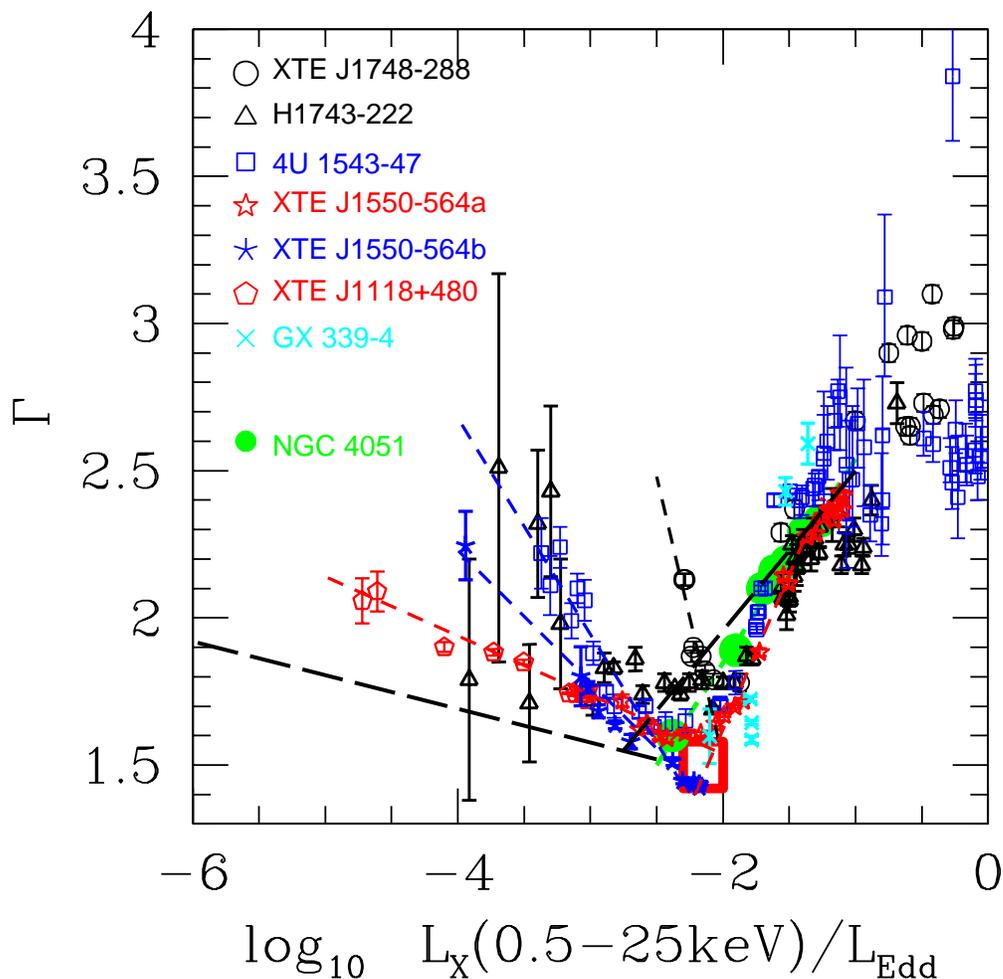} \caption{ The relation between the
hard X-ray photon index and the Eddington ratio in XRBs. The short
dashed lines are linear least-square fits on the anti-correlation
and the positive correlation (see text for details). The long
dashed-lines are the linear least-square fits on LLAGNs (left, Gu \&
Cao 2008) and QSOs (right, Shemmer et al. 2006). The green solid
circles are seven epoch observations of NGC 4051. The red square
represents the range of crosspoints of the anti-correlations and
positive correlation for all XRBs. \label{fig1}}
\end{figure}

\begin{figure}
\epsscale{1.0} \plotone{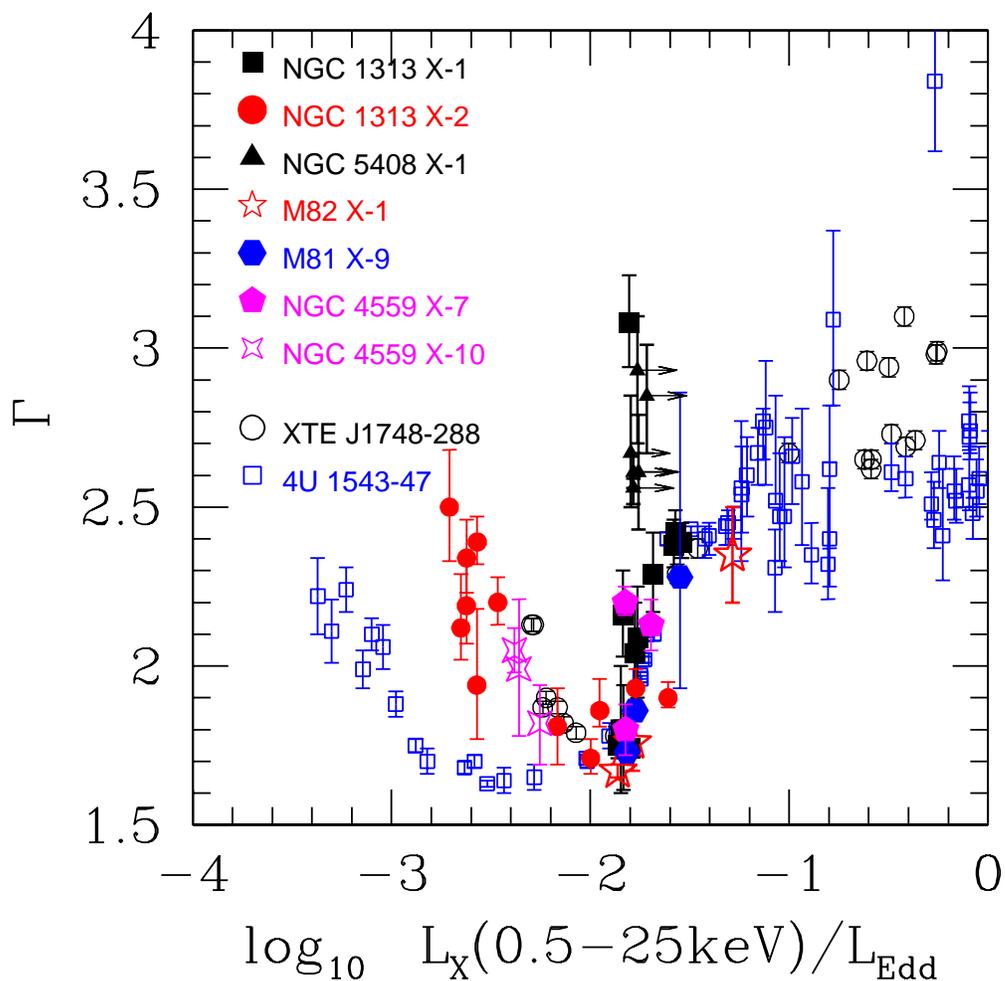} \caption{ The relation between the
hard X-ray photon index and the Eddington ratio in ULXs. The
Eddington luminosity $L_{\rm Edd}$ is calculated from the estimated
BH mass from X-ray spectral evolution in this work. For comparison,
two XRBs (XTE J1748-288 and 4U 1543-47) are also
plotted.\label{fig2}}
\end{figure}

\end{document}